\documentclass[12pt]{article}
\usepackage[english,german,french,polish]{babel}
\usepackage[T1]{fontenc}

\begin{document}

\baselineskip 0.75cm
\topmargin -0.4in
\oddsidemargin -0.1in

\let\ni=\noindent

\renewcommand{\thefootnote}{\fnsymbol{footnote}}

\newcommand{\CKM}{Cabibbo-Kobayashi-Maskawa }

\newcommand{\SM}{Standard Model }

\pagestyle {plain}

\setcounter{page}{1}

\pagestyle{empty}




~~~~~
\begin{flushright}
IFT-02/33
\end{flushright}

\vspace{0.2cm}

{\large\centerline{\bf Explicit seesaw model and deformed fermion universality{\footnote {Work supported in part by the Polish State Committee for Scientific Research (KBN), grant 5 P03B 119 20 (2001--2002).}}}}

\vspace{0.5cm}

{\centerline {\sc Wojciech Kr\'{o}likowski}}

\vspace{0.23cm}

{\centerline {\it Institute of Theoretical Physics, Warsaw University}}

{\centerline {\it Ho\.{z}a 69,~~PL--00--681 Warszawa, ~Poland}}

\vspace{0.5cm}

{\centerline{\bf Abstract}}

\vspace{0.3cm}

In the simple model of neutrino texture presented in this paper, the Majorana left\-handed mass matrix is zero, the Majorana righthanded mass matrix --- diagonal and degenerate, and the Dirac mass matrix has a hierarchical structure, deformed {\it unitarily} by nearly bimaximal mixing. In the case, when the Majorana righthanded term dominates over the Dirac term, the familiar seesaw mechanism leads effectively to the nearly bimaximal oscillations of active neutrinos, consistent with solar and atmospheric neutrino experiments. If the Dirac term, before its unitary deformation, is {\it similar} in shape to the known charged-lepton mass matrix, then parameters for solar $\nu_e$'s and atmospheric $\nu_\mu$'s become related to each other, {\it predicting} from the SuperKamiokande value of $\Delta m_{32}^2$ a tiny $\Delta m_{21}^2$ typical for MSW LOW solar solution {\it rather} than for MSW Large Mixing Angle solution. The predicted mass spectrum is then hierarchical. In Appendix a suggestive form of nearly bimaximal effective mass matrix is derived.

\vspace{0.4cm}

\ni PACS numbers: 12.15.Ff , 14.60.Pq , 12.15.Hh .

\vspace{0.8cm}

\ni August 2002

\vfill\eject

~~~~~
\pagestyle {plain}

\setcounter{page}{1}

\vspace{0.3cm}
 
\ni {\bf 1. Introduction.} The popular, nearly bimaximal form of mixing matrix for three active neutrinos $\nu_{e L}$, $\nu_{\mu L}$, $\nu_{\tau L}$ [1],

\begin{equation} 
U =   \left( \begin{array}{ccc} c_{12} & s_{12} & 0 \\ -s_{12}c_{23} & c_{12}c_{23} & s_{23} \\ s_{12}s_{23} & -c_{12}s_{23} & c_{23} \end{array} \right) \; ,
\end{equation}

\vspace{0.2cm} 

\ni arises from its generic shape {\it \`{a} la} \CKM [2] by putting $s_{13} = 0$ and $c_{12}\,,\, s_{12}\,,\, c_{23}\,,\, s_{23}$ not so far from $1/\sqrt{2}$ ($c_{ij} = \cos \theta_{ij}$ and $s_{ij} = \sin \theta_{ij}$). As is well known, this form is globally consistent with neutrino oscillation experiments [3] for solar $\nu_e$'s and atmospheric $\nu_\mu$'s as well as with the negative Chooz experiment for reactor $\bar{\nu}_e$'s. It cannot explain, however, the possible LSND effect for accelerator $\bar{\nu}_\mu$'s that, if confirmed by the MiniBooNE experiment, may require the existence of one, at least, extra (sterile) light neutrino $\nu_{s L}$ (different, in general, from the conventional sterile neutrinos $(\nu_{e R})^c\,,\, (\nu_{\mu R})^c\,,\, (\nu_{\tau R})^c$). 

The neutrino mixing matrix $U = \left(U_{\alpha i}\right)$ defines the unitary transformation

\begin{equation} 
\nu_{\alpha L}  = \sum_i U_{\alpha i}  \nu_{i L} \;
\end{equation} 

\ni between the active-neutrino flavor and mass fields,  $\nu_{\alpha L}\;(\alpha = e, \mu, \tau)$ and $\nu_{i L}\;(i = 1,2,3)$, respectively. In the flavor representation, where the charged-lepton mass matrix is diagonal, it is at the same time the diagonalizing matrix for the neutrino mass matrix $M = \left(M_{\alpha \beta}\right)$,

\begin{equation} 
U^{\dagger} M U = {\rm diag}(m_1 \,,\,m_2 \,,\,m_3)\,,
\end{equation} 

\ni where $m_1 \,,\,m_2 \,,\,m_3$ denote neutrino masses (real numbers). Two possible Majorana phases in $M$ are assumed to be zero. Then, 

\begin{equation} 
M =  U \, {\rm diag}(m_1\,,\,m_2\,,\,m_3)\, U^\dagger \,,
\end{equation} 

\ni leading in the case of form (1) of $U$ to the following mass matrix elements:

\vfill\eject

\begin{eqnarray} 
M_{e e} & = & \;\,m_1 c^2_{12}+ m_2 s_{12}^2 \;, \nonumber \\
M_{\mu \mu} & = &\; (m_1 s^2_{12}+ m_2 c_{12}^2) c_{23}^2 + m_3 s^2_{23}\;, \nonumber \\
M_{\tau \tau} & = &\; (m_1 s^2_{12}+ m_2 c_{12}^2) s_{23}^2 + m_3 c^2_{23} \;, \nonumber \\ 
M_{e \mu} & = & \!\!\! -(m_1 - m_2) c_{12}s_{12} c_{23}\;\; = M_{\mu e}\;, \nonumber \\
M_{e \tau} & = & \; (m_1 - m_2) c_{12}s_{12} s_{23}\;\; = M_{\tau e}\;, \nonumber \\ 
M_{\mu \tau} & = & \!\!\!- (m_1s^2_{12}+ m_2c_{12}^2  - m_3) c_{23}s_{23} = M_{\tau \mu}\;. 
\end{eqnarray} 

\ni Here, $M^* = M$ and $M^T = M$.

 For the nearly bimaximal form (1) of $U$ the following popular neutrino oscillation probabilities hold (in the vacuum):

\begin{eqnarray} 
P(\nu_e \rightarrow \nu_e)_{\rm sol}\;\;\;\, & = & 1 - (2c_{12}s_{12})^2 \sin^2 (x _{21})_{\rm sol}\;, \nonumber \\
P(\nu_\mu \rightarrow \nu_\mu)_{\rm atm} \;\;& = & 1 - (2c_{23}s_{23})^2 \left[ s^2_{12}\sin^2 (x _{31})_{\rm atm} + c^2_{12}\sin^2 (x _{32})_{\rm atm}\right] \nonumber \\
 & \simeq & 1 - (2c_{23}s_{23})^2 \sin^2 (x _{32})_{\rm atm} \;, \nonumber \\
P(\bar{\nu}_\mu \rightarrow \bar{\nu}_e)_{\rm LSND} & = & (2c_{12}s_{12})^2 c_{23}^2 \sin^2 (x _{21})_{\rm LSND} \simeq 0 \;, \nonumber \\ 
P(\bar{\nu}_e \rightarrow \bar{\nu}_e)_{\rm Chooz} & = & 1 - (2c_{12}s_{12})^2 \sin^2 (x _{21})_{\rm Chooz} \simeq 1\;, 
\end{eqnarray} 

\ni where $\Delta m^2_{21} \ll \Delta m^2_{32} \simeq \Delta m^2_{31}$ and

\begin{equation} 
x_{ji} = 1.27 \frac{\Delta m^2_{ji} L}{E} \;,\; \Delta m^2_{ji}  = m^2_j - m^2_i \;(i,j = 1,2,3)
\end{equation} 

\ni ($\Delta m^2_{ji}$, $L$ and $E$ are measured in eV$^2$, km and GeV, respectively).  Here, $U^* = U$ and $U^T = U^\dagger = U^{-1}$, thus the possible CP violation in neutrino oscillations is ignored. The fourth of these formulae is consistent with the negative Chooz experiment and the third excludes the LSND effect. 

Experimental estimations for solar $\nu_e$'s and atmospheric $\nu_\mu$'s, based on the MSW versions of the first and second formulae (6), are $\theta_{12} \sim (33^\circ \;{\rm or}\; 40^\circ)$, $|\Delta m^2_{21}| \sim (5.5\times 10^{-5}$ or $7.3\times 10^{-8})\;{\rm eV}^2$ [4] and $\theta_{23} \sim 45^\circ$, $|\Delta m^2_{32}| \sim 2.7\times 10^{-3}\;{\rm eV}^2$ [5], respectively. For solar $\nu_e$'s they correspond to the MSW Large Mixing Angle solution or MSW LOW solution, respectively, of which the first is favored. The mixing angles $\theta_{12}$ and $\theta_{23}$ give $c_{12} \sim (1.2/\sqrt{2}$ or $1.1/\sqrt{2})$,  $s_{12} \sim (0.77/\sqrt{2}$ or $0.91/\sqrt{2})$ and $c_{23} \sim 1/\sqrt{2} \sim s_{23}$. The mass-squared differences are  hierarchical, $\Delta m^2_{21} \ll \Delta m^2_{32} \simeq \Delta m^2_{31}$, while the mass spectrum may be {\it either} hierarchical, $ m^2_1 < m^2_2 \ll m^2_3$ with $\Delta m^2_{32} \simeq  m^2_3$, {\it or} nearly degenerate, $ m^2_1 \simeq m^2_2 \simeq m^2_3$ with $\Delta m^2_{21} \ll m^2_2$ and $\Delta m^2_{32} \ll m^2_3$ (here, the ordering $ m^2_1 \leq m^2_2 \leq m^2_3$ is used). If $m_1 \rightarrow 0$, then the option of hierarchical spectrum is true [in spite of nearly bimaximal neutrino mixing expressed by Eq. (1)]. The possibility of $m_1 \rightarrow 0$ is suggested in Section 4.

The rate of neutrinoless double $\beta$ decay (allowed only in the case of  Majorana-type $\nu_{e L} $) is proportional to $m^2_{e e}$, where $m_{e e} \equiv |\sum_i U^2_{e i}m_i| = c^2_{12} |m_1| + s^2_{12} |m_2| \sim (0.70 |m_1| + 0.30 |m_2|$ or $0.59 |m_1| + 0.41 |m_2|)$ and so $|m_1| \leq m_{e e} \leq  |m_2|$ (in our argument $U_{e 3} = 0$ exactly). The suggested experimental upper limit for $m_{e e}$ is $m_{e e} \stackrel{<}{\sim} (0.35 - 1)$ eV [6]. If the actual $m_{e e}$ lay near its upper limit, then the option of nearly degenerate spectrum (with hierarchical mass-squared differences) would be suggested.

\vspace{0.3cm}

\ni {\bf 2. Four-parameter nearly bimaximal texture.} In consistency with the SuperKamiokande data [5] we can put $c_{23} = 1/\sqrt{2} = s_{23}$. Then, in the neutrino mixing and mass matrices, (1) and (4), there are only four independent parameters $m_1\,,\,m_2\,,\,m_3$ and $s_{12}$ leading through Eqs. (5) to four independent matrix elements

\begin{eqnarray} 
M_{e e} & = & \; m_1 c^2_{12}+ m_2 s_{12}^2 \;, \nonumber \\
M_{\mu \mu} & = &\; M_{\tau \tau} = \frac{1}{2}(m_1 s^2_{12}+ m_2 c_{12}^2 + m_3)\;, \nonumber \\
M_{e \mu} & = & - M_{e \tau} = -\frac{1}{\sqrt2}(m_1 - m_2) c_{12}s_{12}\;, \nonumber \\
M_{\mu \tau} & = & - \frac{1}{2}(m_1s^2_{12}+ m_2c_{12}^2  - m_3) \;.
\end{eqnarray} 

\ni Hence, $M_{e e} + M_{\mu \mu} - M_{\mu \tau} = m_1 + m_2$, $M_{e e} - M_{\mu \mu} + M_{\mu \tau} = (m_1 - m_2)(c^2_{12} - s^2_{12})$ and  $(M_{e e} - M_{\mu \mu} + M_{\mu \tau})^2 + 8M^2_{e \mu}  = (m_1 - m_2)^2$.

In this case, the neutrino mass spectrum and mixing can be expressed by four independent parameters [7,8]. Taking as the independent parameters the matrix elements (8) we obtain exactly [7]: 

\begin{eqnarray} 
 m_{1,2}\!\!\! & = & \frac{m_1+m_2}{2}\mp \frac{|m_1-m_2|}{2} = \frac{M_{e e}\!+\! M_{\mu \mu}\! -\!  M_{\mu \tau}}{2} \mp \sqrt{\left(\frac{ M_{e e} - M_{\mu \mu}\!+ \!M_{\mu \tau}}{2}\right)^2\! + \!2 M^2_{e \mu} }, \nonumber \\
m_3 & = &  M_{\mu \mu} + M_{\mu \tau}\,,
\end{eqnarray} 

\ni if $m_2 - m_1 \geq 0$ (both for positive or negative $m_1$ and $m_2$). Signs $\mp$ are replaced here by $\pm$, if $m_1 - m_2 \geq 0$. For the mixing angle $\theta_{12}$ we get

\begin{equation} 
\sin^2 2\theta_{12}= (2 c_{12}s_{12})^2  = \frac{ 8M_{e \mu}^2}{(M_{e e} - M_{\mu \mu} + M_{\mu \tau})^2 + 8M_{e \mu}^2}\;,
\end{equation} 

\ni where $\sin 2\theta_{12} > 0$ if $c_{12}s_{12} > 0$. The formulae (9) and (10) provide us with an inversion of Eqs. (8). At the end of Section 5 we come back to these formulae.

\vspace{0.3cm}

\ni {\bf 3. Explicit seesaw.} Assume now that $M$ is the effective neutrino Majorana mass matrix for active neutrinos, arising by means of the familiar seesaw mechanism [9] from the generic $6\times 6$ neutrino mass term

\begin{equation} 
- {\cal L}_{\rm mass} = \frac{1}{2} \sum_{\alpha \beta} \left( \overline{(\nu_{\alpha L})^c} \,,\, \overline{\nu_{\alpha R}}\right) \left( \begin{array}{cc} M^{(L)}_{\alpha \beta} & M^{(D)}_{\alpha \beta} \\ M^{(D)}_{\beta \alpha} & M^{(R)}_{\alpha \beta} \end{array} \right) \left( \begin{array}{c} \nu_{\beta L} \\ (\nu_{\beta R})^c \end{array} \right) + {\rm h.\,c.} 
\end{equation} 

\ni including both the active neutrinos $\nu_{\alpha L}$ and $(\nu_{\alpha L})^c$ as well as the (conventional) sterile neutrinos $\nu_{\alpha R}$ and $(\nu_{\alpha R})^c \;(\alpha = e\,,\,\mu\,,\,\tau)$. In the seesaw case, the Majo\-rana righthanded mass matrix $ M^{(R)} = \left(M^{(R)}_{\alpha \beta}\right)$ is presumed to dominate over the Dirac  mass matrix $ M^{(D)} = \left(M^{(D)}_{\alpha \beta}\right)$ that in turn  dominates over the Majorana lefthanded mass matrix $ M^{(L)} = \left(M^{(L)}_{\alpha \beta}\right)$ which is expected naturally to be zero (as violating the electroweak gauge symmetry in a nonrenormalizable way in the doublet Higgs case). Then, in the seesaw approximation

\begin{equation} 
M = -M^{(D)} M^{(R) -1}M^{(D)T}\;.
\end{equation} 

\ni Hence, through Eq. (4) we infer that

\begin{equation} 
- M^{(D)} M^{(R) -1}M^{(D)T} = U\;{\rm diag}(m_1,m_2,m_3) U^\dagger
\end{equation}

\ni with $U$ as given in Eq. (1).

The seesaw formula (13) gets an explicit realization in the simple model of neutrino texture, where we postulate that [10]

\begin{equation} 
M^{(L)} = 0 \,,\, M^{(D)} = U\;{\rm diag}(\lambda_1,\lambda_2,\lambda_3) U^\dagger \,,\, M^{(R)} = \mp \Lambda \;{\rm diag}(1,1,1)
\end{equation} 

\ni and then infer that

\begin{equation} 
m_i = \pm \frac{\lambda^2_i}{\Lambda}\;\;\;(i = 1,2,3)\;,
\end{equation} 

\ni respectively, $\lambda_i$ and $\Lambda$ being massdimensional parameters, such that $ 0 \leq \lambda_i \ll \Lambda$. Thus, in this model, $M^{(D)}$ is a unitary transform (through the nearly bimaximal mixing matrix $U$) of the diagonal, potentially hierarchical matrix ${\rm diag}(\lambda_1, \lambda_2, \lambda_3)$, while $M^{(R)}$ is a diagonal, degenerate matrix. In a slightly more general model, $M^{(R)}$ may be also a unitary transform (through the same $U$) of the diagonal, nearly degenerate matrix ${\rm diag}(\Lambda_1, \Lambda_2, \Lambda_3)$ with $\Lambda_1 \simeq \Lambda_2 \simeq \Lambda_3$ (what is natural for large $\Lambda_i$); then $m_i = \pm \lambda^2_i /\Lambda_i$, where $ 0 \leq \lambda_i \ll \Lambda_i\;\,(i = 1,2,3)$. From the first Eq. (14) and Eq. (1) the Dirac mass matrix $M^{(D)} = \left( M^{(D)}_{\alpha \beta} \right)$ gets analogical entries as those given in Eqs. (5) for $M = (M_{\alpha \beta})$, but with $m_i$ replaced now by $\lambda_i = \sqrt{\Lambda |m_i|}\;\;(i = 1,2,3)$, where also Eqs. (15) are used.

We should like to stress that, in the present paper [in particular, in its part pertaining to the simple neutrino model defined through Eq. (14)], the nearly bimaximal mixing matrix $U$ given in Eq. (1) is adopted phenomenologically on the ground of neutrino oscillation data, and so, is by no means derived theoretically. One may speculate that, perhaps, such a derivation would require some new, additional concepts about the nature of connections between neutrinos and charged leptons. The idea of deformed fermion universality presented in the next Section (and, in our opinion, very natural) does not help to explain the experimental appearance of nearly bimaximal neutrino mixing, though this idea coexists nicely with such a mixing (in spite of the hierarchical neutrino mass spectrum implied by it), since the form (5) of $M$ is consistent with $U$ given in Eq. (1) for {\it any} mass spectrum. However, if the form of $M$ could be accepted as natural because of some theoretical reasons, then the nearly bimaximal mixing matrix $U$ implied by such an $M$ might also be considered as justified theoretically. In Appendix, we rewrite $M$, given as in Eq. (8) with $ c_{23} = 1/\sqrt2 = s_{23}$, in a suggestive form that may help to accept it as the neutrino effective mass matrix.  

\vfill\eject


\ni {\bf 4. Deformed fermion universality.}We find very natural the idea that in the first Eq. (14) the original Dirac mass matrix ${\rm diag}(\lambda_1, \lambda_2, \lambda_3)$, before it gets its actual form $M^{(D)}$ deformed unitarily by the nearly bimaximal mixing matrix $U$, is {\it similar} in shape to the charged-lepton and quark mass matrices which are also of the Dirac type. To proceed a bit further with this idea we will try to conjecture that ${\rm diag}(\lambda_1, \lambda_2, \lambda_3)$ has a shape analoguous to the following charged-lepton mass matrix [11]:

\begin{equation}  
{M}^{(e)} = \frac{1}{29} \left(\begin{array}{ccc} \mu^{(e)}\varepsilon^{(e)} & 0 & 0 
\\ 0 & 4\mu^{(e)}(80 + \varepsilon^{(e)})/9 & 0 
\\ 0 & 0 & 24\mu^{(e)} (624 + \varepsilon^{(e)})/25 \end{array}\right)  
\end {equation}  

\ni which predicts accurately the mass $m_\tau = M^{(e)}_{\tau \tau}$ from the experimental values of masses $m_e = M^{(e)}_{e e}$ and $m_\mu = M^{(e)}_{\mu \mu}$ treated as an input. In fact, we get $m_\tau = 1776.80$ MeV [11] {\it versus} $m^{\rm exp}_\tau = 1777.03^{+0.30}_{-0.26}$ MeV [12] and, in  addition, determine $\mu^{(e)} = 85.9924$ MeV and $\varepsilon^{(e)} = 0.172329$. For a theoretical background of this particular form of $M^{(e)}$ the interested reader may consult Ref. [13]. Let us emphasize that the figures in the mass matrix (16) are not fitted {\it ad usum Delphini}.

Thus, making use of the neutrino analogue $M^{(\nu)}$ of $M^{(e)}$ given in Eq. (16), we put ${\rm diag}(\lambda_1, \lambda_2, \lambda_3) = M^{(\nu)}$.Then [10],

\begin{eqnarray}
\sqrt{\Lambda |m_1|} & = & \lambda_1 =  \frac{\mu^{(\nu)}}{29} \,\varepsilon^{(\nu)} = 0.0345\,\varepsilon^{(\nu)} \,\mu^{(\nu)}  \;, \nonumber \\
\sqrt{\Lambda |m_2|} & = & \lambda_2 =  \frac{\mu^{(\nu)}}{29}\,\frac{4(80+\varepsilon^{(\nu)})}{9}  = \left(1.23 + 0.0153\,\varepsilon^{(\nu)}\right) \,\mu^{(\nu)}  \;, \nonumber \\
\sqrt{\Lambda |m_3|} & = & \lambda_3 =  \frac{\mu^{(\nu)}}{29}\frac{24(624+\varepsilon^{(\nu)})}{25} = \left(20.7+ 0.0331\,\varepsilon^{(\nu)}\right)\mu^{(\nu)}  \;, 
\end{eqnarray}

\ni where also Eqs. (15) are invoked. Hence, taking $\varepsilon^{(\nu)} = 0$ (already $\varepsilon^{(e)}$ is small), we calculate

\begin{equation}  
m^2_1 = 0 \;,\;m^2_2 = 2.26 \frac{\mu^{(\nu)\,4}}{\Lambda^2} \;,\;m^2_3 = 1.82\times 10^5 \frac{\mu^{(\nu)\,4}}{\Lambda^2} 
\end {equation}  

\ni and

\vspace{-0.2cm}

\begin{eqnarray}  
\Delta m^2_{21} = m^2_2 = 2.26 \frac{\mu^{(\nu)\,4}}{\Lambda^2}\!\! & , &\!\!  \Delta m^2_{32} = m^2_3 - m^2_2  = 1.82\times 10^5 \frac{\mu^{(\nu)\,4}}{\Lambda^2} \;, \nonumber \\
\Delta m^2_{21} /\Delta m^2_{32}\!\!  & = &\!\!  1.24\times 10^{-5}\;.
\end{eqnarray}   

\ni The neutrino mass spectrum described by Eqs. (18) is hierarchical, $ m^2_1 < m^2_2 \ll m^2_3$, in spite of the appearance of nearly bimaximal neutrino mixing. Using in the second Eq. (19) the SuperKamiokande estimate $\Delta m^2_{32}  \sim 2.7\times 10^{-3}\;{\rm MeV}^2$ [5], we get

\begin{equation}  
\mu^{(\nu)\,2} \sim 1.2 \times 10^{-4}\Lambda \;{\rm eV}\;.
\end {equation}  

\ni If taking reasonably $\mu^{(\nu)} \,\stackrel{<}{\sim} \mu^{(e)} = 85.9924$ MeV, we obtain from Eq. (20) that $ \Lambda\,\stackrel{<}{\sim} \, 6.1 \times 10^{10}$ GeV. If $\varepsilon^{(\nu)}\,\leq \varepsilon^{(e)} = 0.172329$ ({\it i.e}, not necessarily $\varepsilon^{(\nu)} =0$), then $m^2_{1}/m^2_{3} \leq 6.81\times 10^{-15}$ from Eq. (17) and so, with $m^2_3 \sim 2.7\times 10^{-3}\;{\rm eV}^2$ we estimate $m_1^2 \stackrel{<}{\sim} 1.8\times 10^{-17}\;{\rm eV}^2$, thus $m^2_1 = 0$ practically.

\vspace{0.3cm}

\ni {\bf 5. Conclusions.} From the ratio $\Delta m^2_{21}/ \Delta m^2_{32} $ in Eq. (19) and the estimate $ \Delta m^2_{32} \sim 2.7\times 10^{-3}\;{\rm eV}^2$ we obtain the {\it prediction} [10]

\begin{equation}  
m^2_2 = \Delta m^2_{21} \sim 3.3\times 10^{-8}\;{\rm eV}^2
\end {equation}

\ni which lies not so far from the experimental estimate $ \Delta m^2_{21} \sim 7.3\times 10^{-8}\;{\rm eV}^2$ based on the MSW LOW solar solution [4], whereas the favored experimental estimation based on the MSW Large Mixing Angle solar solution is much larger: $ \Delta m^2_{21} \sim 5.5 \times 10^{-5}\;{\rm eV}^2$. So, if really true, the latter excludes drastically our conjecture (17). Otherwise, this conjecture might be a significant step  forwards in our understanding of neutrino texture, in particular, of the question of fermion universality extended to neutrinos.

If the prediction $m^2_1 = 0$, $m^2_2 \sim 3.3\times 10^{-8}\;{\rm eV}^2$ and $m^2_3 \sim 2.7\times 10^{-3}\;{\rm eV}^2$ were true, then our previous estimate $m_{e e} \sim 0.59|m_1| + 0.41|m_2|$ of the effective mass of $\nu_e$ in the neutrinoless double $\beta$ decay would give $m_{e e} \sim 7.5\times 10^{-5}$ eV, dramatically below the presently suggested experimental upper limit $m_{e e} \stackrel{<}{\sim}$ (0.35 -- 1) eV [6] (recall, however, that in our argument $ U_{e 3} =0$ exactly). In this case, the option of hierarchical mass spectrum, $m^2_1 < m^2_2 \ll m^2_3$, would be true. This would be true also for $m^2_2 \simeq \Delta m^2_{21} \sim 5.5 \times 10^{-5}\;{\rm eV}^2$. 

When $m_1 = 0$ (as in the case of our conjecture (17) with $\varepsilon^{(\nu)} = 0$), the four parameters in the mass formula (9) can be related to three independent parameters $M_{e e}$, $M_{\mu \mu}$ and $M_{\mu \tau}$, since in this case the first Eq. (9) gives

\begin{equation}  
M^2_{e \mu} = \frac{1}{2} M_{e e} (M_{\mu \mu} -  M_{\mu \tau}).
\end {equation}

Then, from the second and third Eq. (9) we obtain

\begin{eqnarray}  
m_{2} & = &  M_{e e} + M_{\mu \mu} -  M_{\mu \tau} \sim \pm 1.8\times 10^{-4}\,{\rm eV}
\;, \nonumber \\
m_{3} & = &  M_{\mu \mu} +  M_{\mu \tau} \sim \pm 5.2\times 10^{-2}\,{\rm eV}\;,
\end{eqnarray}

\ni where we use also the estimates $m_2^2 \sim 3.3\times 10^{-8}{\rm eV}^2$ and $m_3^2 \sim 2.7\times 10^{-3}{\rm eV}^2$. Here, as already in Eqs (9) and (15), we allow for positive or negative neutrino masses. Similarly, when $m_1 = 0$, the formula (10) gives

\begin{equation} 
\sin^2 2\theta_{12} = \frac{ 4M_{e e}(M_{\mu \mu} - M_{\mu \tau})}{(M_{e e} + M_{\mu \mu} - M_{\mu \tau})^2} \sim (0.84\;\,{\rm or}\;\,0.97) 
\end{equation} 

\ni due to Eq. (22), where also the estimate $\theta_{12} \sim (33^\circ \;\,{\rm or}\;\,40^\circ)$ is used. Hence, $ s^2_{12} = M_{e e}/(M_{e e} + M_{\mu \mu} - M_{\mu \tau})$ and $ c^2_{12} = (M_{\mu \mu } - M_{\mu \tau})/(M_{e e} + M_{\mu \mu} - M_{\mu \tau})$.

We can see from Eqs (15) and (23) that for $m_1 = 0$

\begin{eqnarray}
\lambda^2_1 & = & 0 \;, \nonumber \\ \lambda^2_2  & = &  \pm \Lambda (M_{e e} + M_{\mu \mu} - M_{\mu \tau}) \sim 1.8\times 10^{-4} \Lambda \;\,{\rm eV}\;, \nonumber \\
\lambda^2_3  & = & \Lambda (M_{\mu \mu} + M_{\mu \tau}) \sim 5.2\times 10^{-2} \Lambda \;\,{\rm eV}\;,
\end{eqnarray}

\ni where $\lambda_1,\lambda_2,\lambda_3$ and $\Lambda $ are mass parameters introduced in Eqs. (14). In our simple neutrino model defined through Eqs. (14), where $M^{(R)}$ is diagonal and degenerate, the formulae (25) express for $m_1 = 0$ the seesaw relationship $M^{(D)}M^{(D)T} = - M^{(R)} M$, equivalent to Eq. (12) (as $M^{(R)-1}$ and $M^{(D)}$ commute).

Finally, let us mention that if, instead of the model of neutrino texture defined in Eqs. (14), we had [14]

\begin{equation} 
M^{(L)} = \pm \Lambda\; {\rm diag}(1,1,1) \,,\, M^{(D)} = U\;{\rm diag}(\lambda_1,\lambda_2,\lambda_3) U^\dagger \,,\, M^{(R)} = 0 \,,
\end{equation} 

\ni then under the assumption of $0\leq \lambda_i \ll \Lambda$ we would obtain for active neutrinos the effective mass matrix of the form

\begin{equation} 
M = M^{(L)} + M^{(D)} M^{(L) -1} M^{(D) T} = U\;{\rm diag}(\lambda_1,\lambda_2,\lambda_3) U^\dagger 
\end{equation} 

\ni with the nearly degenerate mass spectrum $ m_i = \pm(\Lambda + \lambda^2_i/\Lambda)\;(i = 1,2,3)$. Here, $\Lambda \gg \lambda_i$, but much less dramatically than in the seesaw mechanism working in Eqs. (14). In this case, when making the conjecture of deformed fermion universality as it is expressed in Eqs. (17), we would predict $\Delta m^2_{21}$ of the order $10^{-5}\;{\rm eV}^2$, not very far from the favored experimental estimate $5.5\times 10^{-5}\;{\rm eV}^2$ based on the MSW Large Mixing Angle solar solution (now, $m^{(\nu) 2} \sim 3.2\times 10^{-6}\;{\rm eV}^2$). The nonzero $M^{(L)}$ given in Eqs. (26) would not be justified, however, in the doublet Higgs case, since it would violate the electroweak gauge symmetry in a nonrenormalizable way (though, possibly, spontaneously). In the case of the model (26), oscillations between and into the (conventional) sterile neutrinos $(\nu_{e R})^c\,,\, (\nu_{\mu R})^c\,,\, (\nu_{\tau R})^c$ would be negligible as $(\lambda_i/\Lambda)^2$. For these sterile neutrinos the effective mass matrix would be $-M^{(D)} M^{(L) -1} M^{(D)T}\!$, implying the mass spectrum $\mp \lambda^2_i/\Lambda $.

\vspace{0.5cm}

{\centerline{\bf Appendix}} 

\vspace{0.1cm}

{\centerline{\it A suggestive form of effective neutrino mass matrix}} 

\vspace{0.3cm}

The neutrino spectrum (9), valid when $c_{23} = 1/\sqrt2 = s_{23}$, can be rewritten in the case of $0 \leq m_1 < m_2$ in the form

\vspace{-0.2cm}

$$
m_{1,2} = \,\stackrel{0}{m}\, \mp \,\delta \;\;,\;\; m_3 = \,\stackrel{0}{m}\, + \Delta \;, 
\eqno{\rm (A.1)}
$$

\vspace{-0.3cm}

\ni where

\vspace{-0.4cm}

\begin{eqnarray*}
0 < \,\stackrel{0}{m}\,  & = & \frac{M_{e e} + M_{\mu \mu} - M_{\mu \tau}}{2}\;, \\
0 < \delta\; & = & \sqrt{\left(\frac{M_{e e} - M_{\mu \mu} + M_{\mu \tau}}{2}\right)^2 + 2M^2_{e \mu}}\;, \\
 \Delta & = & M_{\mu \mu} + M_{\mu \tau} - \,\stackrel{0}{m}\,  = \frac{- M_{e e} + M_{\mu \mu} + 3M_{\mu \tau}}{2}\;. 
\end{eqnarray*}

\vspace{-1.55cm}

\begin{flushright}
({\rm A}.2)
\end{flushright}

\vspace{-0.3cm}

\ni Then, Eqs. (8) give

\begin{eqnarray*}
M_{e e} & = & \stackrel{0}{m}\, - \;\delta \cos 2\theta_{12} \;,\\
M_{\mu \mu} & = & M_{\tau \tau} = \;\stackrel{0}{m}\, + \frac{1}{2}\Delta + \frac{1}{2}\delta \cos 2\theta_{12} \;,\\
M_{e \mu} & = & \!\!- M_{e \tau} = \frac{1}{\sqrt2}\,\delta \sin 2\theta_{12} \;,\\
M_{\mu \tau} & = & \!\!- \frac{1}{2}\Delta - \frac{1}{2}\delta \cos 2\theta_{12}  \;.
\end{eqnarray*}

\vspace{-1.6cm}

\begin{flushright}
({\rm A}.3)
\end{flushright}

\ni Thus, with the use of Eqs. (A.3), the effective mass matrix $M = \left(M_{\alpha \beta} \right)$ may be preseneted as follows:

$$
M = \,\stackrel{0}{m}\, \left(\begin{array}{ccc} 1 & 0 & 0 \\ 0 & 1 & 0 \\ 0 & 0 & 1\end{array}\right) \!+\! \Delta \left(\begin{array}{ccc} 0 & 0 & 0 \\ 0 & \frac{1}{2} & \frac{1}{2} \\ 0 & \frac{1}{2} & \frac{1}{2}\end{array}\right) \!+\! \delta \left(\begin{array}{ccc} -\cos 2\theta_{12} & \frac{1}{\sqrt2} \sin 2\theta_{12} & -\frac{1}{\sqrt2} \sin 2\theta_{12} \\ \frac{1}{\sqrt2} \sin 2\theta_{12} & \frac{1}{2}\cos 2\theta_{12} & -\frac{1}{2}\cos 2\theta_{12} \\ -\frac{1}{\sqrt2} \sin 2\theta_{12} & -\frac{1}{2}\cos 2\theta_{12} & \frac{1}{2}\cos 2\theta_{12}\end{array}\right)   \;,
\eqno{\rm (A.4)}
$$

\vspace{0.2cm}

\ni where three component matrices commute with each other (the products in two orderings of the second and third matrix vanish). Using the nearly bimaximal mixing matrix $U$ defined in Eq. (1), now in the form

$$
U =   \left( \begin{array}{ccc} c_{12} & s_{12} & 0 \\ - \frac{1}{\sqrt2} s_{12} & \frac{1}{\sqrt2} c_{12} & \frac{1}{\sqrt2}  \\ \frac{1}{\sqrt2} s_{12} & -\frac{1}{\sqrt2} c_{12} & \frac{1}{\sqrt2}  \end{array} \right) 
\eqno{\rm (A.5)}
$$

\vspace{0.2cm}

\ni with $c_{23} = 1/\sqrt2 = s_{23}$, we obtain

$$
\left(\begin{array}{ccc} m_1 & 0 & 0 \\ 0 & m_2 & 0 \\ 0 & 0 & m_3\end{array}\right) \!=\! U^\dagger MU = \,\stackrel{0}{m}\,\left(\begin{array}{ccc} 1 & 0 & 0 \\ 0 & 1 & 0 \\ 0 & 0 & 1\end{array}\right) \!+\! \Delta \left(\begin{array}{ccc} 0 & 0 & 0 \\ 0 & 0 & 0 \\ 0 & 0 & 1\end{array}\right)  \!+\! \delta \left( \begin{array}{ccc} -1 & 0 & 0 \\ 0 & 1 & 0 \\ 0 & 0 & 0\end{array}\right), 
\eqno{\rm (A.6)}
$$

\vspace{0.2cm}

\ni consistently with Eq. (A.1). Naturally, the mass matrix $M$ determines its diagonalizing matrix $U$ and the mass spectrum $m_1, m_2, m_3$ (here, the mixing matrix is at the same time the diagonalizing matrix).

The form (A.4) of the effective neutrino mass matrix clearly suggests the full democracy of $\nu_\mu $ and $\nu_\tau $ neutrinos, and of their interactions with $\nu_e$ neutrino. These interactions are described by the third component matrix that becomes

\vspace{0.2cm}

$$
\delta \left( \begin{array}{ccc}   0 & \frac{1}{\sqrt2}  & -\frac{1}{\sqrt2} \\ \frac{1}{\sqrt2} & 0 & 0 \\ -\frac{1}{\sqrt2}  & 0 & 0 \end{array}\right), 
\eqno{\rm (A.7)}
$$

\ni for $c_{12} \rightarrow 1/\sqrt2 \leftarrow s_{12}$. Correspondingly, Eq. (A.5) and the formula $\nu_i = \sum_\alpha U^*_{\alpha i} \nu_\alpha$, inverse to Eq. (2), express the full democracy of $\nu_\mu $ and $\nu_\tau $, and of their mixings with $\nu_e$, leading to 

\vspace{-0.4cm}

\begin{eqnarray*}
\nu_1 & = & c_{12} \nu_e - s_{12} \frac{\nu_\mu - \nu_\tau}{\sqrt2} \;, \\
\nu_2 & = & s_{12} \nu_e + c_{12} \frac{\nu_\mu - \nu_\tau}{\sqrt2} \;, \\
\nu_3 & = & \frac{\nu_\mu + \nu_\tau}{\sqrt2} 
\end{eqnarray*}

\vspace{-1.45cm}

\begin{flushright}
({\rm A}.8)
\end{flushright}

\vspace{-0.3cm}

\ni or

\vspace{-0.5cm}

\begin{eqnarray*}
\nu_1 & = & \frac{1}{\sqrt2} \left( \nu_e -  \frac{\nu_\mu - \nu_\tau}{\sqrt2}\right) \;,\\
\nu_2 & = & \frac{1}{\sqrt2} \left( \nu_e + \frac{\nu_\mu - \nu_\tau}{\sqrt2}\right) \;,\\
\nu_3 & = & \frac{\nu_\mu + \nu_\tau}{\sqrt2} 
\end{eqnarray*}

\vspace{-1.45cm}

\begin{flushright}
({\rm A}.9)
\end{flushright}

\ni for $c_{12} \rightarrow 1/\sqrt2 \leftarrow s_{12}$. In this limit there are two maximal mixings: $\nu_\mu$ with $\nu_\tau$ into the superpositions $(\nu_\mu \mp \nu_\tau)/\sqrt2$, and $\nu_e$ with $(\nu_\mu - \nu_\tau)/\sqrt2$ into $[\nu_e \mp (\nu_\mu - \nu_\tau)/\sqrt2]/\sqrt2$. The above interpretation of $M$ given in Eq. (A.4) is independent of the values of $\stackrel{0}{m},\,\delta$ and $\Delta$. 

Making use of the mass-squared differences

$$
\Delta m^2_{21} = m^2_2  - m^2_1 = 4\stackrel{0}{m}\delta \;,\; \Delta m^2_{32} = m^2_3 - m^2_2  = 2 \stackrel{0}{m} (\Delta - \delta) + \Delta^2 - \delta^2
\eqno({\rm A}.10)
$$

\ni following from Eq. (A.1), we can write for $\Delta > 0$

$$
\delta = \frac{\Delta m^2_{21}}{4\,\stackrel{0}{m}\,}\;,\; \Delta = -\stackrel{0}{m}\, + \sqrt{(\,\stackrel{0}{m}\, + \delta)^2 + \Delta m^2_{32}}\;.
\eqno({\rm A}.11)
$$

\ni Let us consider two extremal options: ({\it i}) $\delta \simeq \,\stackrel{0}{m}\, < \Delta$, where $\,\stackrel{0}{m}\, \simeq \delta \simeq \sqrt{\Delta m^2_{21}} / 2 $ due to the first Eq. (A.11) and $\Delta = -\stackrel{0}{m}\, + \sqrt{\Delta m^2_{21}+\Delta m^2_{32}}$ from the second and first Eqs. (A.11), and ({\it ii}) $\delta \ll \Delta \ll \,\stackrel{0}{m}\, $, where $ \delta = \Delta m^2_{21} / 4\!\stackrel{0}{m}\, $ and $\Delta \simeq \delta + \Delta m^2_{32}/2\!\stackrel{0}{m}\, $ from Eqs (A.11). Here, $\Delta m^2_{21} \ll \Delta m^2_{32}$ from the experiment.

In the option ({\it i}), taking the experimental estimates $\Delta m^2_{21} \sim (5.5\times 10^{-5}$ or $7.3\times 10^{-8}\;{\rm eV}^2)$ and $\Delta m^2_{32} \sim 2.7\times 10^{-3}\;{\rm eV}^2$, we obtain

$$
\stackrel{0}{m}\, \simeq \delta \simeq \frac{1}{2} \sqrt{\Delta m^2_{21}} \sim (3.7\times 10^{-3}\,{\rm or}\, 1.4\times 10^{-4})\,{\rm eV},\Delta \simeq \sqrt{\Delta m^2_{32}} - \,\stackrel{0}{m}\, \sim (4.8\,{\rm or}\, 5.2)\times 10^{-2}\,{\rm eV}\,
\eqno({\rm A}.12)
$$

\ni and hence, the mass spectrum

$$
m_1 \simeq 0\;,\; m_2 \simeq 2{\stackrel{0}{m}}\, \sim (7.4\times 10^{-3}\;{\rm or}\; 2.7\times 10^{-4})\,{\rm eV}\;,\; m_3 = \,\stackrel{0}{m}\, + \Delta \sim 5.2\times 10^{-2}\,{\rm eV}
\eqno({\rm A}.13)
$$

\ni that is hierarchical, $0 \simeq m^2_1 < m^2_2 \ll m^2_3 $ [here, $m^2_2/m^2_3 \sim (2.0\times 10^{-2}$ or $2.7\times 10^{-5}$)].

In the option ({\it ii}), with the same estimates for $\Delta m^2_{21}$ and $\Delta m^2_{32}$ we get

$$
\delta \simeq \frac{\Delta m^2_{21}}{4\stackrel{0}{m}} \sim (1.4\times 10^{-5}\;{\rm or}\; 1.8\times 10^{-8})\, \frac{{\rm eV}^2}{\stackrel{0}{m}} \;,\; \Delta \simeq \frac{\Delta m^2_{32}}{2\stackrel{0}{m}} \sim 1.4\times 10^{-3} \frac{{\rm eV}^2}{\,\stackrel{0}{m}\,} 
\eqno({\rm A}.14)
$$

\ni and then, the mass spectrum

$$
m_1 \simeq \,\stackrel{0}{m}\,\,,\, m_2 \simeq \,\stackrel{0}{m}\,\,,\, m_3 = \,\stackrel{0}{m} + \Delta \sim \,\stackrel{0}{m}\, + 1.4\times 10^{-3}\,\frac{{\rm eV}^2}{\,\stackrel{0}{m}\,}
\eqno({\rm A}.15)
$$

\ni which is nearly degenerate, $\,\stackrel{0}{m}\, \simeq m_1 \simeq m_2 \simeq m_3$, if $\,\stackrel{0}{m}\, \gg 1.4\times 10^{-3}\;{\rm eV}^2/\,\stackrel{0}{m}\,$ (in this case, also $ \stackrel{0}{m}\!^2\simeq m^2_1 \simeq m^2_2 \simeq m^2_3$, of course). For instance, if $\,\stackrel{0}{m}\, \sim 1$ eV, then $\delta \sim (1.4\times 10^{-5}$ or $1.8\times 10^{-8}$) eV and $\Delta \sim 1.4\times 10^{-3}$ eV, so that $\delta \ll \Delta \ll \,\stackrel{0}{m}\,$.

The conjecture of deformed fermion universality, discussed in Section 4 and concluded in Section 5, may work in the case of option ({\it i}) (though it is not obligatory), but it cannot be applied in the case of option ({\it ii}).

When applying to the option ({\it i}) the conjecture of deformed fermion universality (with $\varepsilon^{(\nu)} = 0$) and making use of the estimate $\Delta m^2_{32} \sim 2.7\times 10^{-3}\;{\rm eV}^2$  leading to $\Delta m^2_{21} \sim 3.3\times 10^{-8}\;{\rm eV}^2$, Eq. (21), we obtain 

$$
\,\stackrel{0}{m}\, = \delta = \frac{1}{2} \sqrt{\Delta m^2_{21}} \sim 0.91\times 10^{-4}\,{\rm eV}\;,\; \Delta \simeq  \sqrt{\Delta m^2_{32}}\, - \;\stackrel{0}{m}\, \sim 5.2\times 10^{-2}\,{\rm eV}
\eqno({\rm A}.16)
$$

\ni in place of Eq. (A.12). This new value for $\,\stackrel{0}{m}\, = \delta$ is a {\it prediction} lying not so far from the former value $1.4\times 10^{-4}$ eV of $\,\stackrel{0}{m}\, = \delta$ following from the estimation $\Delta m^2_{21} \sim 7.3\times 10^{-8}\;{\rm eV}^2$ based on the MSW LOW solar solution. Then,

$$
m_1 = 0\,,\, m_2 = 2{\stackrel{0}{m}}\, \sim 1.8\times 10^{-4}\;{\rm eV}\,,\, m_3 = \,\stackrel{0}{m} + \Delta \sim 5.2\times 10^{-2}\;{\rm eV}
\eqno({\rm A}.17)
$$

\ni in place of Eq. (A.13).

\vspace{0.2cm}

\vfill\eject

~~~~
\vspace{0.5cm}

{\centerline{\bf References}}

\vspace{0.45cm}

{\everypar={\hangindent=0.6truecm}
\parindent=0pt\frenchspacing

{\everypar={\hangindent=0.6truecm}
\parindent=0pt\frenchspacing

[1]~~{\it Cf. e.g.} Z. Xing, {\it Phys. Rev.} {\bf D 61}, 057301 (2000); and references therein.

\vspace{0.2cm}

[2]~~Z. Maki, M. Nakagawa and S. Sakata, {\it Prog. Theor. Phys.} {\bf 28}, 870 (1962).

\vspace{0.2cm}

[3]~~For a recent review {\it cf.} M.C. Gonzalez-Garcia and Y.~Nir, {\tt hep--ph/0202056}; and references therein.

\vspace{0.2cm}

~[4]~V. Barger, D. Marfatia, K. Whisnant and B.P.~Wood, {\tt hep--ph/0204253}; J.N.~Bahcall, M.C.~Gonzalez--Garcia and C. Pe\~{n}a--Garay, {\tt hep--ph/0204314v2}; G.L.~Fogli {\it et al.}, {\tt hep-ph/0208026}; and references therein.

\vspace{0.2cm}

[5]~~S. Fukuda {\it et al.}, {\it Phys. Rev. Lett.} {\bf 85}, 3999 (2000).

\vspace{0.2cm}

[6]~~{\it Cf. e.g.} M. Frigerio and A.Yu. Smirnov, {\tt hep--ph/0202247}; and references therein.

\vspace{0.2cm}

[7]~~W. Kr\'{o}likowski, {\tt hep-ph/0007255}. 

\vspace{0.2cm}

[8]~~E.~Ma, {\tt hep--ph/0207352}; {\tt hep--ph/0208077}.

\vspace{0.2cm}

[9]~~M. Gell-Mann, P. Ramond and R.~Slansky, in {\it Supergravity}, edited by F.~van Nieuwenhuizen and D.~Freedman, North Holland, 1979; T.~Yanagida, Proc. of the {\it Workshop on Unified Theory and the Baryon Number in the Universe}, KEK, Japan, 1979; R.N.~Mohapatra and G.~Senjanovi\'{c}, {\it Phys. Rev. Lett.} {\bf 44}, 912 (1980).

\vspace{0.2cm}

[10]~{\it Cf.} W. Kr\'{o}likowski, {\tt hep-ph/0207086v2}.

\vspace{0.2cm}

[11]~W. Kr\'{o}likowski, {\it Acta Phys. Pol.} {\bf B 27}, 2121 (1996); and references therein. 

\vspace{0.2cm}

[12]~The Particle Data Group, {\it Eur. Phys. J.} {\bf C 15}, 1 (2000).

\vspace{0.2cm}

[13]~W. Kr\'{o}likowski, Appendix in {\it Acta Phys. Pol.} {\bf B 32}, 2961 (2001); Appendix B in 
{\it Acta Phys. Pol.} {\bf B 33}, 1747 (2002); {\tt hep--ph/0203107}; and references therein.

\vspace{0.2cm}

[14]~{\it Cf.} W. Kr\'{o}likowski, {\it Acta Phys. Pol.} {\bf B 33}, 641 (2002). 

\vspace{0.2cm}

\vfill\eject

\end{document}